\documentclass{article}

\pdfoutput=1
\usepackage{authblk}
\usepackage[margin=1in]{geometry}
\usepackage[onehalfspacing]{setspace}
\usepackage[utf8]{inputenc}
\usepackage{graphicx}
\usepackage{booktabs}
\usepackage{tikz}
\usepackage{amsmath}
\usepackage{amsfonts}
\usepackage{color, colortbl}
\usepackage[colorlinks = true,
            linkcolor = blue,
            urlcolor  = blue,
            citecolor = blue]{hyperref}

\usepackage[mathlines]{lineno}
\usepackage[round]{natbib}
\usepackage[most]{tcolorbox}

\usepackage[T1]{fontenc}
\usepackage[utf8]{inputenc}
\usepackage{mathptmx}

\usepackage[mathscr]{eucal}

\title{Emergence as the conversion of information: A unifying theory}
\author[1,2]{Thomas Varley}
\author[3]{Erik Hoel\thanks{hoelerik@gmail.com}}
\affil[1]{Department of Psychological \& Brain Sciences, Indiana University, Bloomington, IN, USA}
\affil[2]{School of Informatics, Computing, and Engineering, Indiana University, Bloomington, IN, USA}
\affil[3]{Allen Discovery Center, Tufts University, Medford, MA, USA}

\begin{document}

\maketitle

\begin{abstract}

Is reduction always a good scientific strategy? Does it always lead to a gain in information? The very existence of the special sciences above and beyond physics seems to hint no. Previous research has shown that dimension reduction (macroscales) can increase the dependency between elements of a system (a phenomenon called "causal emergence"). However, this has been shown only for specific measures like effective information or integrated information. Here, we provide an umbrella mathematical framework for emergence based on information conversion. Specifically, we show evidence that a macroscale can have more of a certain type of information than its underlying microscale. This is because macroscales can convert information from one type to another. In such cases, reduction to a microscale means the loss of this type of information. We demonstrate this using the well-understood mutual information measure applied to Boolean networks. By using the partial information decomposition, the mutual information can be decomposed into redundant, unique, and synergistic information atoms. Then by introducing a novel measure of the synergy bias of a given decomposition, we are able to show that the synergy component of a Boolean network's mutual information can increase at macroscales. This can occur even when there is no difference in the total mutual information between a macroscale and its underlying microscale, proving information conversion. We relate this broad framework to previous work, compare it to other theories, and argue it complexifies any notion of universal reduction in the sciences, since such reduction would likely lead to a loss of synergistic information in scientific models.


\end{abstract}

\section{\label{sec:introduction}Introduction}

Reductionism is one of the classic principles of science. At the same time, science itself forms a diverse tree with elements at different spatiotemporal scales, such as quantum waves in physics, molecules in chemistry, cells in biology, all the way up to macroeconomics and sociology. Macroscale descriptions like biophysical models of cells, the machine code in computers, or organisms operating within a food web, are generally treated as if they reflect some intrinsic scale of function that cannot be neatly improved by reduction. The result is a contradiction between the theory and practice of science \citep{fodor1974special, hoel2018agent}. 

One resolution to this contradiction is the "null hypothesis" of reductionism: that all macroscale descriptions, which broadly are some form of dimension reduction such as coarse-graining, are only useful due to computational constraints. This is because, according to this hypothesis, their underlying microscales contain all the information. That is, \textit{information compression} is the only true benefit to analyzing, modeling, or understanding a system at a macroscopic level. Compression of a given information source can be lossless or lossy, but can never lead to an overall gain of information \citep{cover2012elements}. Without any gain of information at a macroscale some have argued that macroscales cannot add anything above or beyond their underlying microscales and therefore should be considered epiphenomenal \citep{kim1998mind, bontly2002supervenience}.

An alternative resolution to the contradiction between universal reductionism and science as practiced is a formal theory of emergence. Such a formal theory should a) directly and fairly compare microscales to macroscales, b) offer a quantitative measurement of what a macroscale is providing in terms of information gain above and beyond compression, and c) enable the means to identify emergent scales in a given system or data set. A non-trivial formal theory of emergence should allow for reduction or emergence on a case-by-case basis. Such a theory of emergence can solve longstanding problems in model choice for scientists, since it reveals the intrinsic scale of function of systems. 

The first contemporary formal theory of emergence took the form of "causal emergence" \citep{hoel2013quantifying}. It made use of the effective information ($EI$), which is the mutual information between a set of interventions by an experimenter and their effects. More specifically, the effective information is the mutual information following an experimenter intervening to set a system or part of a system to maximum entropy. Since the interventions cannot have any common causes, any correlations of future system states to that injected noise must be caused by that noise. Furthermore, the effective information has been shown mathematically to capture the causally-relevant information in a system by being sensitive to the determinism (lack of uncertainty in state transitions) and degeneracy (similar or identical state transitions or dynamics, e.g., the necessity of a given state transition). Dimension reductions like coarse-graining (grouping elements of states in macro-elements or macro-states) or black-boxing (leaving elements or states exogenous) can increase the effective information \citep{hoel2017map}. Overall, the argument was that since macro-states were more deterministic and less degenerate, they constrained the past/future of the system to a greater degree, and the effective information was able to identify the spatiotemporal scale at which this constraint peaked. This peak indicated that this scale, whether macro or micro, was the most causally-relevant scale.

One possible criticism of the theory is that the effective information is only one specific measure of causation. Originally proposed by Giulio Tononi and Olaf Sporns in 2003 \citep{tononi2003measuring}, effective information is mathematically well-described \citep{balduzzi2008integrated, balduzzi2011information, hoel2013quantifying}. While it keeps being re-invented as a measure of causation \citep{korb2009information, griffiths2015measuring}, generally without acknowledgement of previous formulations or ongoing lines of research, the measure has not yet been proven to be the unique measure of causation, and there are alternative proposals (many of which are mathematically related) for how to measure causation using information theory \citep{albantakis2019caused}. Indeed, the same general phenomena of causal emergence has been shown in integrated information \citep{hoel2016can, marshall2018black} as the $\phi$ measure can increase at macroscales due to similar reasons of increasing determinism and decreasing degeneracy of state transitions. Measures that are not directly causal but capture aspects of causation, like assessing the entropy of random walkers on networks (indicating uncertainty of transitions), can also improve at macroscales in that random walkers are more deterministic in their dynamics \citep{klein2020emergence}. This all indicates there is a broader phenomenon at work. Specifically, there is somehow more causally-relevant information at macroscales (although it is currently unclear if this is captured by a unique measure of causation, or is better captured by a set of common measurements). How is such information gain possible?

Information cannot be created \textit{ex nihilo}. Here we propose that emergence is a form of \textit{information conversion} at a higher scales. When measured in its totality in a given system, total information measures like the entropy of the distribution system states, the Kolmogorov information for describing the entirety of the system, or the total correlation in the form of the mutual information, all necessarily decrease at a macroscale (or at best, do not decrease, but can never increase). However, information can be converted from one type to another, with no change except for what scale the system is being modeled at, meaning there can be a gain of specific types of information at macroscales. 

Herein we demonstrate this phenomenon of information conversation making use of the classic and well-understood measure of mutual information. Specifically, we consider the mutual information between the past and future of Boolean networks \citep{kauffman1969homeostasis}. While total mutual information always decreases or remains constant at a macroscale (no information \textit{ex nihilo}), this is not the full story, since the information itself can be decomposed into a set of partial information ``atoms" (PI atoms) that quantify how the total information is distributed over all of the elements of the system \citep{williams_nonnegative_2010}. Herein we show that, after coarse-graining, redundant information at the microscale can be converted into synergistic information at macroscales in an overall movement of information up the PI lattice, and that this effect exists even when no mutual information is lost. This indicates that the ``structure" of the mutual information is truly being converted from redundant to synergistic at macroscales.

In Section \ref{sec:MI_PID}, we overview the how we are applying the mutual information in our model system, and also its decomposition in multi-element systems. We introduce a novel measure of redundancy/synergy bias in the mutual information, based on how PI-atoms are distributed across the PI-lattice. In Section \ref{sec:macroscales},
we examine systems of Boolean networks across scales and accompanying changes in the partial information decomposition. First we look at common macroscales of logic gates and find a clear shift toward synergies in some systems (e.g., we show how an XOR is more synergistic than its underlying microscale logic gates, none of which are XORs). Next, we examine sets of Boolean networks wherein the mutual information is identical at both micro and macroscales, and show that even in these cases there can be an increase in synergy bias at the macroscale. This shows direct evidence of information conversion at macroscales in that the mutual information becomes more dominated by its synergistic component without changing its total bit value, just its decomposition. In Section \ref{sec:CE_as_information_conversion}, we connect our work to previous research by demonstrating how causal emergence can be thought of as a form of information conversion, as the total entropy of transitions is converted to the causally-relevant form of effective information via dimension reduction. Overall, we conclude that information conversion offers an agnostic umbrella explanation for theories of emergence based in information theory.

\section{\label{sec:MI_PID}Mutual information and its decomposition in discrete systems}

Here we detail our application of the mutual information to probabilistic Boolean networks. Systems can be viewed as passing information from the past to the future over the channel of the present. The quantification of this information flow can be done by calculating the mutual information between the future and past joint states of the variables that make up a network. Specifically, $X$ represents the past states of the Boolean network, while $Y$ represents the future states of the network. The calculation of $I(X,Y)$ quantifies how much knowledge of the past state of the system reduces our uncertainty about the future state of the system. Specifically:

\begin{equation}
\label{eq:mi}
    I(X ; Y) = \sum_{x\in\mathcal{X}}\sum_{y\in\mathcal{Y}} P(x, y) \log_2\big(\frac{P(x,y)}{P(x)P(y)}\big)
\end{equation}

This calculation requires defining the distributions $P(X)$, $P(Y)$, and $P(X,Y)$. The joint distribution is given by the transition probability matrix (TPM) of the system (with each row weighted by the probability of that state), and $P(X)$ is an ``input" distribution. To not bias our measurements, $P(X)$ is the stationary distribution of the system (in cases of multiple stationary distributions we use the one with the largest attractor, or design Boolean networks such that all states are included in a single attractor). The ``output" distribution ($P(Y)$) can then be calculated as the matrix-multiplication of $P(X)$ and $P(X,Y)$. Note that every ``state" in the support sets $\mathcal{X}$ and $\mathcal{Y}$ actually represents the joint state of multiple variables in the underlying Boolean network.

This application of the mutual information captures the total amount of information in the dynamics of the system (the calculation of which requires a the system's stationary dynamics $P(X)$). For example, in networks where the stationary distribution contains only a single state in the form of a point attractor the mutual information is zero, since the system is like a source that sends only a single message over a channel: there is no uncertainty about the future to be reduced by knowledge of the past. In networks where each state is visited equally in the stationary distribution, and each state deterministically transitions to a unique state, the mutual information would be maximized as $log_2(n)$, as every "message" the system sends is as informative as possible. 

\subsection{Partial information decomposition}
A core limitation of mutual information when assessing systems with more than two variables is that it gives little direct insight into \textit{how} information is distributed over sets of multiple interacting variables. Consider the classic case of two elements $X_1$ and $X_2$ that regulate a third variable $Y$: it is easy to determine the information shared between either $X_i$ and $Y$ as $I(X_i ; Y)$, and it is possible calculate the joint mutual information $I(X_1, X_2 ; Y)$, however these measures leave it ambiguous what information is associated with which combination of variables. For example, if $X_1 \not{\bot} X_2$, then there is necessarily some information about $Y$ that is redundantly shared between both $X_1$ and $X_2$. Similarly, it is possible that there is \textit{synergistic} information about $Y$ that is only disclosed by the joint states of $X_1$ and $X_2$ together and not retrievable from either variable independently (for example if all elements are binary and $Y=X_1 \oplus X_2$, then $I(X_1 ; Y) = I(X_2 ; Y) = 0 \text{ bit}$ but $I(X_1, X_2 ; Y) = 1 \text{ bit}$). 

To address this issue, the Partial Information Decomposition (PID) framework was introduced \citep{williams_nonnegative_2010}. It provides a method by which the mutual information between the joint state of multiple sources variables and a single target variable can itself be decomposed. For the example case detailed above, with two ``source" variables ($X_1$ and $X_2$) and a single ``target" variable $Y$, the full partial information decomposition breaks $I(X_1, X_2 ; Y)$ down into the following additive combination of ``partial information atoms":

\begin{equation}
\label{eq_pid1}
    I(X_1, X_2 ; Y) = Red(X_1, X_2 ; Y) + Unq(X_1 ; Y | X_2) + Unq(X_2 ; Y | X_1) + Syn(X_1, X_2 ; Y).
\end{equation}

Where $Red(X_1, X_2 ; Y)$ is the information about $Y$ that is \textit{redundantly shared} between $X_1$ and $X_2$ (i.e. an observer could learn the same information about $Y$ examining either $X_1$ or $X_2$), $Unq(X_1 ; Y | X_2)$ refers to the information about $Y$ that is uniquely present in $X_1$ and not in $X_2$, and $Syn(X_1, X_2 ; Y)$ is the information about $Y$ that is \textit{only} disclosed by the joint states of $X_1$ and $X_2$ considered together. Furthermore, the bivariate mutual informations can also be decomposed:

\begin{equation}
\label{eq_pid2}
    I(X_1 ; Y) = Red(X_1, X_2 ; Y) + Unq(X_1 ; Y | X_2)
\end{equation}
\begin{equation}
\label{eq_pid3}
    I(X_2 ; Y) = Red(X_1, X_2 ; Y) + Unq(X_2 ; Y | X_1)
\end{equation}

The result is that Eqs. \ref{eq_pid1}, \ref{eq_pid2}, and \ref{eq_pid3} form an under-determined system of three equations with four unknowns ($Red$, $Unq_1$, $Unq_2$, $Syn$). Given an appropriate function with which to compute any of these three, the rest are trivial.

As the number of sources informing about a single target grows, the number of combinations of sources that must be considered grows super-exponentially. The seminal contribution of Williams and Beer was to realize that it is not necessary to brute-force search every combination in the power-set of sources, but rather, that meaningful combinations of sources are naturally structured into a partially-ordered lattice, known as the partial-information (PI) lattice. Furthermore, for a particular set of  sources $\alpha$, the value of the associated partial information atom $\Pi_\alpha$ can be calculated recursively as the difference between the information redundantly shared across the sources of interest, and the sum of all PI-atoms lower down on the lattice:

\begin{equation}
\label{eq:mobius}
\Pi(\alpha, Y) = Red(\alpha, Y) - \sum_{\beta \prec \alpha}\Pi(\beta, Y)
\end{equation}

Where $Red(\alpha, Y)$ is the \textit{redundancy function}, which quantifies the information about $Y$ that is redundantly present in every element of $\alpha$. 
For readers interested in the deeper mathematical details of the construction of the PI lattice, we refer to \citep{williams_nonnegative_2010}, and more recently \citep{gutknecht_bits_2020} for a more in-depth discussion. For our purposes, it suffices to understand that there exists a partial ordering of PI atoms, with ``more redundant" atoms towards the bottom. For example, in the case of three sources, the bottom of the PI-lattice is given by $\{0\}\{1\}\{2\}$, which refers to the information about the target that is redundantly present in all three sources. At the top of the lattice is $\{012\}$, which gives the information about the target that is only accessible when considering the joint state of all three sources jointly, and not disclosed by any ``simpler" combination of sources. It is important to note that, for systems with more than two sources, the PI-atoms no-longer break down into neatly intuitive categories of ``redundant", ``unique", and ``synergistic": more exotic combinations of sources appear, for example: $\{0\}\{12\}$, which gives the information about the target that is redundantly present in $X_0$ and the joint state of $X_1$ and $X_2$ together. However, in general the lower down on the PI-lattice a PI-atom is, the more redundant the information is, while the higher on the lattice, the more synergistic. 

While the PID framework provides the scaffold on which information can be decomposed, it fails to provide the specific keystone necessary to actually calculate it: the redundancy function that forms the base of the PI-lattice. Williams and Beer proposed the \textit{specific information} as a plausible redundancy function typically denoted as $I_{WB}$:

\begin{equation}
    I_{WB}(\textbf{X} ; Y) = \sum_{y\in\mathcal{Y}}P(y)\min_{X_i \in \textbf{X}}I(X_i ; y)    
\end{equation}

The specific information quantifies the average minimal amount every element of \textbf{X} discloses about \textit{Y}. The term $\min_{X_i \in \textbf{X}}I(X_i ; y)$ calculates the minimal amount of information any $X_i \in \textbf{X}$ provides about the specific state $Y=y$. Across all $y \in \mathcal{Y}$, $I_{WB}$ quantifies the expected minimum amount of information that \textbf{X} will disclose about $Y$. As a redundancy function, $I_{WB}$ has a number of appealing quantities: in contrast to other redundancy functions, it will only return values greater than or equal to zero bit. Given the perennial difficulties of interpreting negatively-valued information quantities, this is a key property, one not shared by most other redundancy functions. Furthermore, $I_{WB}$ is both conceptually and computationally simple: being based on the specific information, it is a ``pure" information theoretic measure and does not require leveraging theory or algorithms from fields like information geometry \citep{harder_bivariate_2013}, game theory \citep{ay_information_2019}, or decision theory \citep{bertschinger_quantifying_2014,kolchinsky_novel_2019} and is one of the fastest-running functions in the \textit{dit} toolbox. $I_{WB}$ is also arguably the most well-used redundancy function in the scientific literature, having been the function of choice in \citep{timme_synergy_2014,timme_high-degree_2016,faber_computation_2018,sherrill_correlated_2020,tax_partial_2017}. 

Williams and Beer's redundancy function has been critiqued for behaving in an ``unintuitive" manner in some cases \citep{griffith_quantifying_2014}, and does not readily localize the way that the mutual information function does \citep{finn_pointwise_2018}. Since PID was initially proposed, considerable work has gone into developing an ``optimal" redundancy function, resulting in a plethora of measures \citep{bertschinger_shared_2013,harder_bivariate_2013,griffith_intersection_2014,griffith_quantifying_2014,olbrich_information_2015,goodwell_temporal_2017,ince_measuring_2017,finn_pointwise_2018,james_unique_2018}. So far, no single measure has emerged as the accepted ``gold standard", although they share many commonalities. Each function comes with it's own limitations (for example, only being defined for two variables, or occasionally returning negative quantities of partial information, or violating some intuitions about how such a measure ``should" behave), so care is necessary when deciding which one to use. 

In this work, we used the original measure put forth by Williams and Beer, as it was necessary that whatever measure we chose never return negative quantities of information, be applicable to systems with more than two sources, and this measure remains widely used and the most studied. We used the Discrete Information Theory package \citep{james_dit_2018} for the implementation of $I_{WB}$ and all PID calculations.

\subsection{PID of temporal mutual information}

Partial information decomposition is usually applied to situations like those given in the example above, where a set of sources (neurons, perceptrons, etc) synapse onto a single target and is often applied in such cases \citep{tax_partial_2017,faber_computation_2018}. Here we detail our application of the partial information decomposition of the mutual information between the past and the future in Boolean networks.

Consider a Markovian system with two interacting elements that is evolving in time. Following the convention given above, we will say that $\textbf{X}=\{X_1, X_2\}$ indicates the past states of both elements of our system, and $\textbf{Y}=\{Y_1,Y_2\}$ indicates the future states of both elements of our system. We can then adapt the classic PID framework by defining our ``sources" as every $X_i \in \textbf{X}$, and our single target as the joint future state \textbf{Y}. The PID of this two-element dynamical system is then given by:

\begin{equation}
I(\textbf{X},\textbf{Y}) = Red(X_1, X_2 ; \textbf{Y}) + Unq(X_1 ; \textbf{Y} | X_2) + Unq(X_2 ; \textbf{Y} | X_1) + Syn(X_1, X_2 ; \textbf{Y})
\end{equation}

This decomposition details how information about the next joint-state of the system is distributed  \citep{mediano_beyond_2019}.



\subsection{\label{sec:synergy_bias}Introducing synergy and redundancy biases}

\begin{figure}
    \centering
    \includegraphics[scale=0.75]{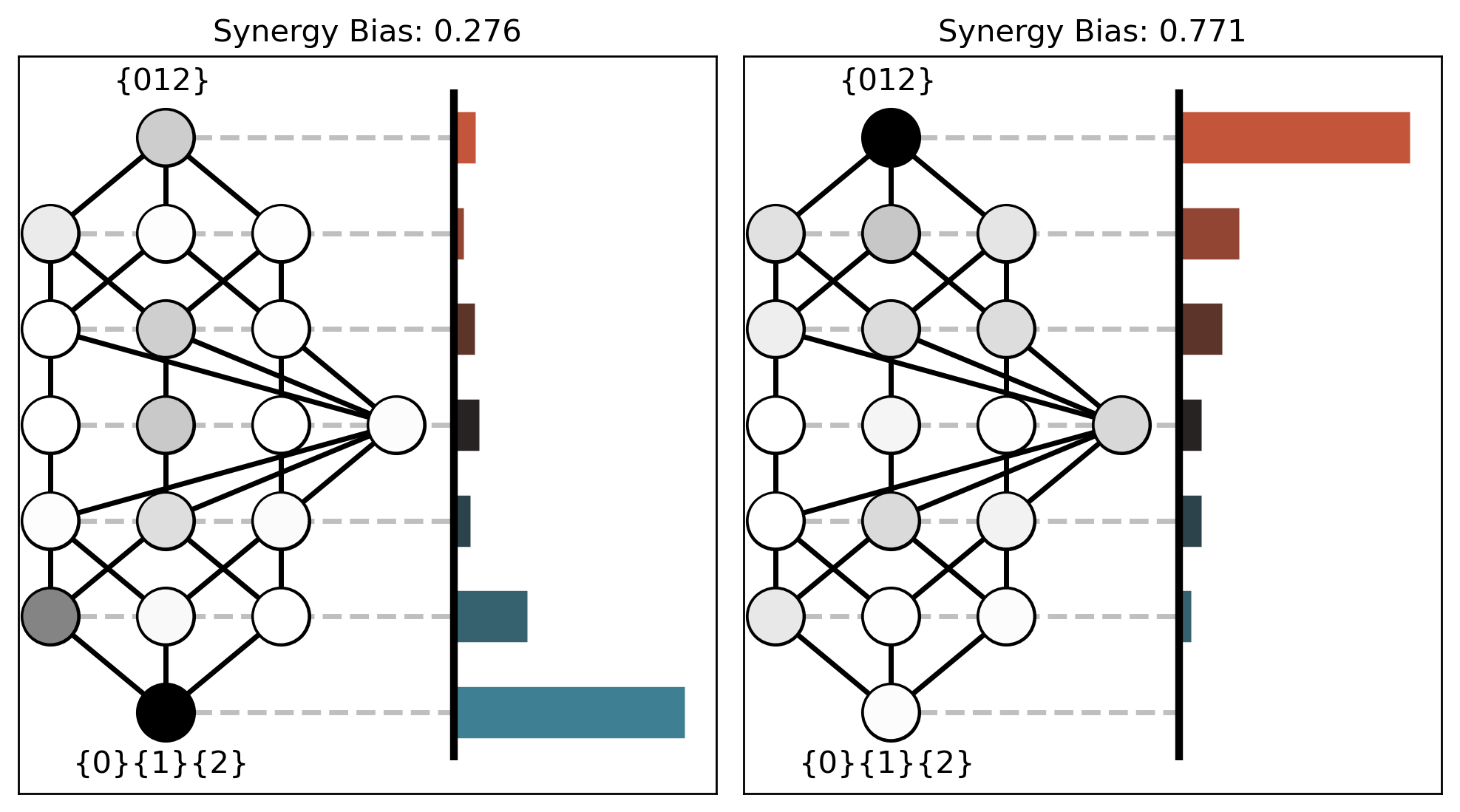}
    \caption{\textbf{The partial information spectrum.} The PI lattices for two, three-element systems. For each system, from the PI lattice we can create a PI spectrum, which gives the proportion of the total mutual information present in all PI atoms at a given ``height" on the lattice. \textbf{Left}: this system has a low synergy-bias (high redundancy bias): the majority of the mutual information about the joint-future state is redundantly shared across all the elements (\{0\}\{1\}\{2\}), or other highly-redundant PI atoms (e.g. \{0\}\{1\}). \textbf{Right}: this 3-element system has a high degree of synergy bias, with the majority of the information about the joint future present only in the joint state of all three elements (\{012\}).
    }
    \label{fig:pi_spectra}
\end{figure}

In our two toy examples above, we relied heavily on the categorical distinction between redundant, unique, and synergistic information. These classifications are useful for building intuition, but do not readily generalize to systems with more than two elements. To address this, we introduce the construct of a \textit{partial information spectrum}, from which one can calculate the relative top- or bottom-heaviness of a PI lattice without directly having to define well-delineated ``pools" of redundant, unique, or synergistic information.


Recall that the value of a given PI atom is calculated recursively from the sums of every PI atom lower than it down on the lattice; PI atoms higher on the lattice contain information that is increasingly synergistic and cannot be extracted from any simpler combination of sources. Because the PI lattice is partially ordered, there are collections of PI atoms that are at the same ``height" on the lattice relative to the bottom (the maximum redundancy atom) or the top (the maximum synergy atom). We claim that these atoms comprise a ``layer" of the lattice defined by some ratio of redundancy to synergy. The PI spectrum $S$ is then defined as an ordered list where the $i^{th}$ bin in the spectrum is given by the proportion of total mutual information present in all PI-atoms in the $i^{th}$ layer.

Once the PI spectrum ($S$) has been calculated, it is easy to determine how top-heavy it is using a measure similar to the Earth Mover's Distance. We define the synergy bias ($B_{syn}(S)$) as the amount of normalized partial information in each layer, weighted by the number of steps that layer is from the bottom.  
\begin{equation}
    B_{syn}(S) = \sum_{i=0}^{|S|} \frac{i}{|S|}S_i
\end{equation}

Where $i$ indexes the layer (starting from bottom, maximally redundant layer) and $|S|$ gives the total number of layers in the lattice. By normalizing by the total number of layers, we can compare the synergy bias between two different sized lattices, since we are looking at the proportion of the total lattice height moved, rather than counting the actual number of layers. 

The redundancy bias is defined equivalently, although the reference point is the top of the lattice, rather than the bottom:

\begin{equation}
    B_{red}(S) = 1 - \sum_{i=0}^{|S|} \frac{i}{|S|}S_i
\end{equation}

Conveniently, $B_{syn} + B_{red} = 1$, so we only ever have to calculate one. 

The synergy and redundancy biases allow us to compare how top- and bottom-heavy two different PI lattices are: a high synergy bias indicates that most of the partial information is present in synergistic relationships between elements, while a high redundancy bias indicates that most of the partial information is redundantly present across multiple individual elements. Since it is a normalized measure, we can compare the top- and bottom-heaviness of systems with different numbers of elements (and consequently, different sized lattices) and thus can measure synergy/redundancy bias across scales.

\section{\label{sec:macroscales}Evidence of information conversion across scales}

\subsection{\label{sec:logic_gates}Macroscales can increase the synergy bias of information}

We begin with a well-known type of macroscale as our model system: that of a single logic gate, which itself is some dimension reduction of a larger collection of networked gates. By breaking down three basic logic gates with distinct mechanisms (AND, OR, and XOR) into collections of microscale gates with simpler mechanisms, we can directly and fairly compare the microscales and macroscales in terms of their respective distribution of partial information. This provides a first showcase of information conversion across scales.

Note that here we are technically leaving elements exogenous in time in our macroscale (since the microscale networks require multiple timesteps to run), and all mechanisms have been coarse-grained into single mechanism (again, broadly these are all referred to as forms of dimension reduction). Such example systems of logic gates have no stationary dynamics, since they are not closed, but are open systems. Therefore, to calculate the mutual information, in all cases (both micro and macro) the same input distribution to either the macroscale mechanism's inputs or the inputs to the network of microscale logic gates that underlies them. For solely explanatory purposes we make use of the maximum entropy as our input distribution in all cases, and this means that $P(X)$ is identical for both macro and microscales in our comparisons.

By calculating the mutual information of the macro and microscales with the same inputs, and decomposing the result using PID, we can see that the synergy bias of the system is not constant between scales. Consider the XOR gate: which can be decomposed into a network of one NAND gate, one AND gate and one OR gate (as well as two inputs A and B). The resulting system has five elements compared to the macroscale, which has 3 elements (XOR, A and B). We found that, at the micro-cale, the system had a mutual information of 2.5 bit, while it had the expected 1 bit of mutual information at the macro scale. However, while the total mutual information decreased for the XOR gate macroscale, the synergy of that information increased, from 0.52 at the microscale to 0.83 at the macroscale. The same was true for both the OR and the AND gates, although to a lesser degree (final values can be found in Table \ref{tab:logic}). Note that while the AND and OR gates have the same macroscale mutual information and synergy bias (since they are isomorphic), they have different microscale values, which reflects the different number and structure of NAND gates required to implement them. 

\begin{figure}
    \centering
    \includegraphics[scale=0.75]{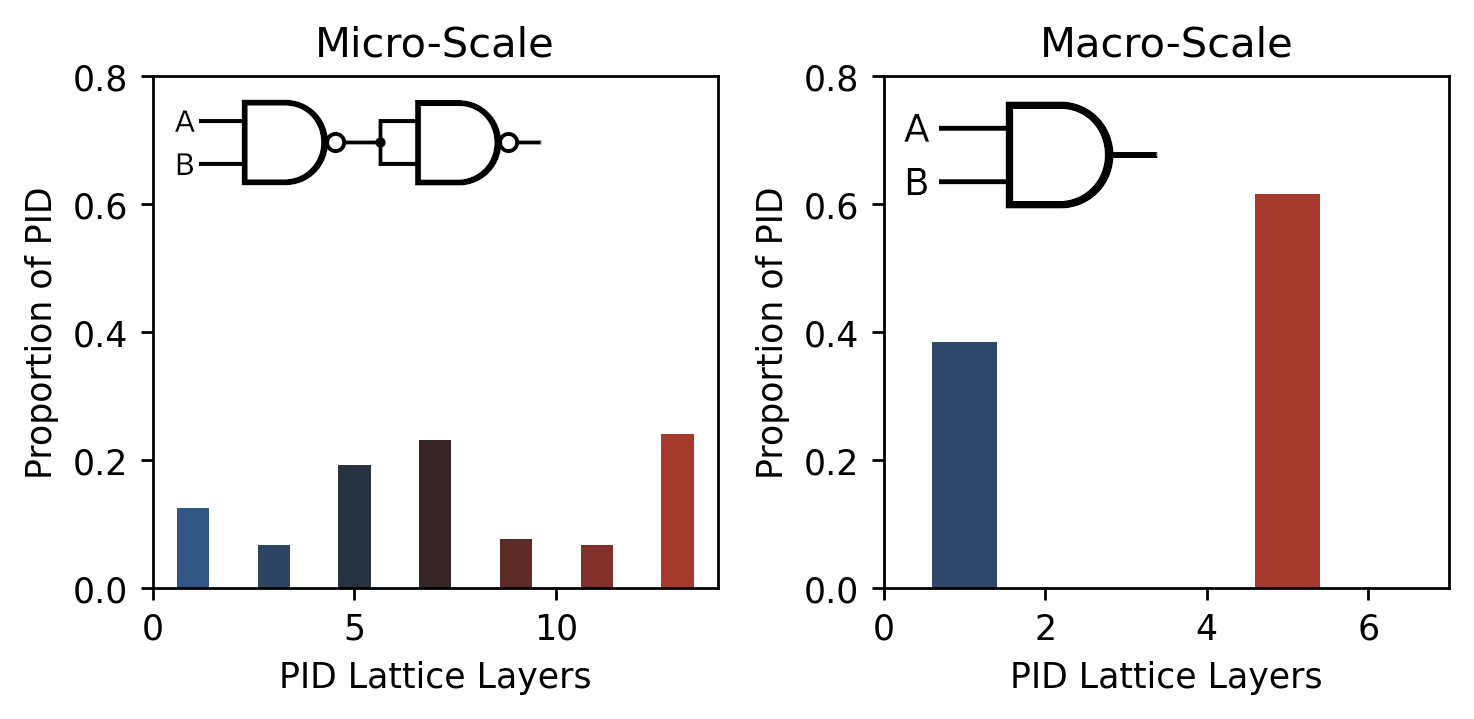}
    \includegraphics[scale=0.75]{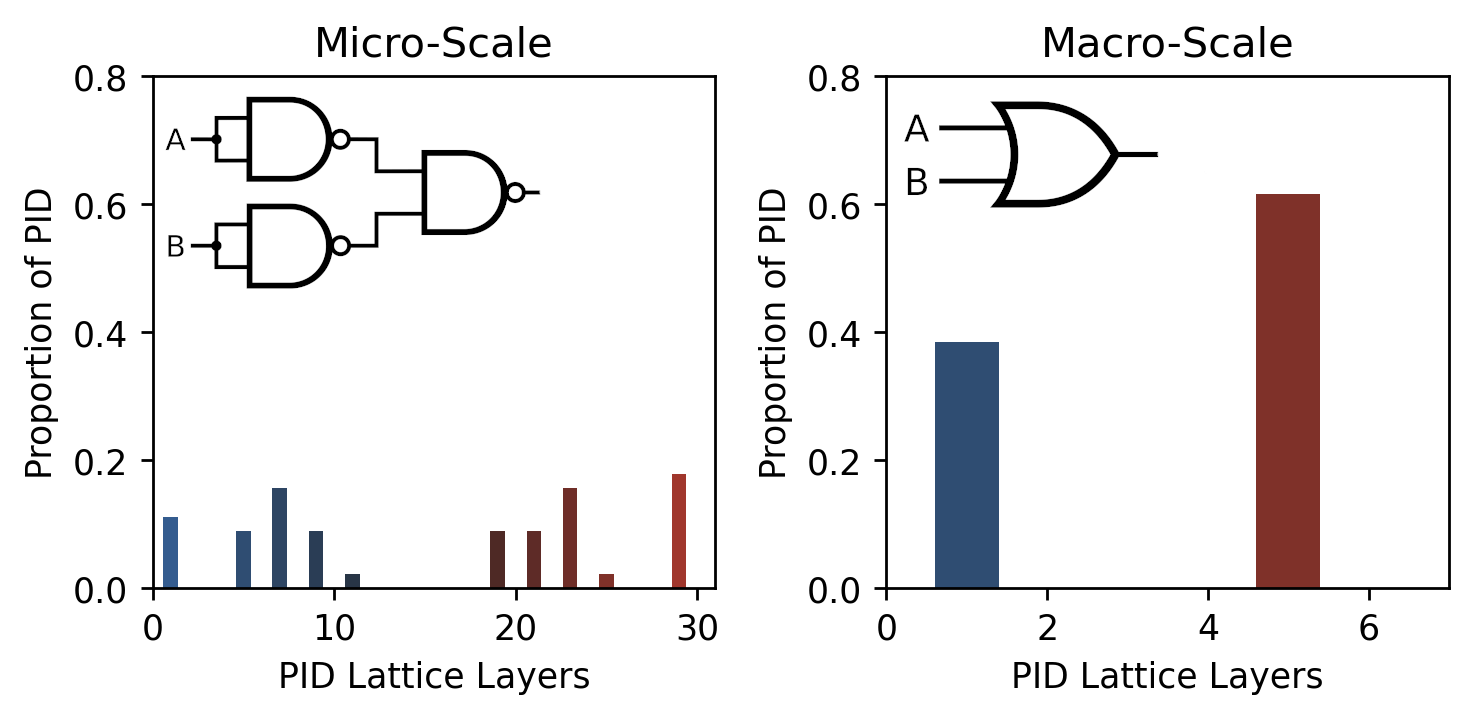}
    \includegraphics[scale=0.75]{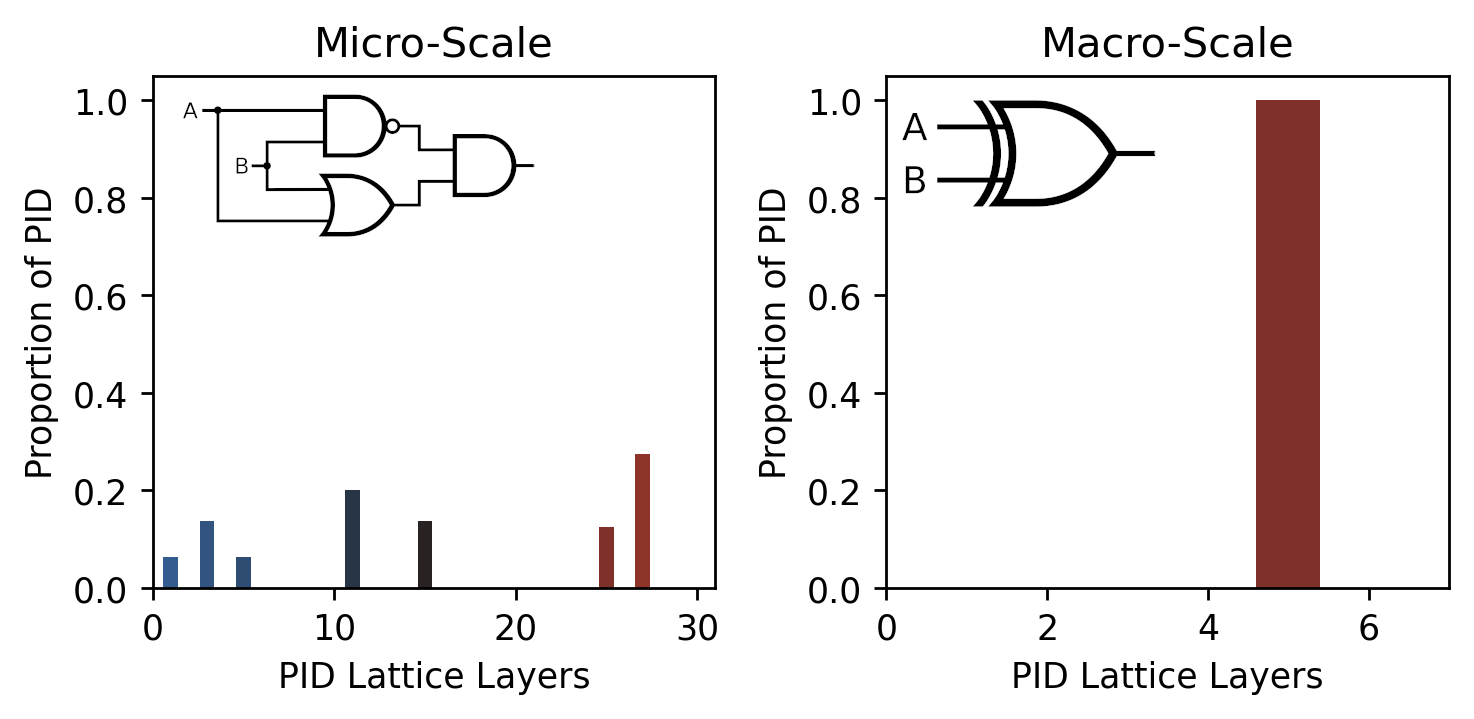}
    \caption{\textbf{Partial-Information Spectra for Logic Gates.} Partial information spectra for three logic gates: AND (\textbf{top}), OR (\textbf{middle}), XOR (\textbf{bottom}) at the macroscale (\textbf{right}) and the microscale (\textbf{left}). The synergy biases and temporal mutual information values can be viewed in Table \ref{tab:logic}.}
    \label{fig:logic_gates}
\end{figure}

This suggests that, while dimension reduction reduces the overall amount of information in a system, the "leftover" information can move higher on the macroscale PID lattice. This can be seen directly in Fig. \ref{fig:logic_gates}, which shows the PI spectra based off of the PI lattices for three macroscale mechanisms and their underlying microscales of networked logic gates.

When considering our logic gate result, it is clear that dimension reductions like coarse-grainings can alter the distribution of information in the PI lattice of a system, even when both scales are simply a different description of the same system. Note that this result fits intuitively with the idea that something is being gained by modeling an XOR gate as an actual XOR gate, even if it is made of a set of underlying logic gates that, like NAND gates, are not themselves XORs. What is gained is a distinctive bias toward synergy in the information flowing through the system. Additionally, it is intuitive that XORs should greatly demonstrate this effect, like ANDs and ORs demonstrate it to only a slight degree. While not shown, it should be obvious that there can be reverse cases; for example, some macroscales may be more redundant than their underlying microscales, since there is no limit to how complex a microscale can be.

\begin{table}[]
\begin{tabular}{@{}lllll@{}}
\toprule
\textbf{Gate} & \textbf{Microscale MI} & \textbf{Macroscale MI} & \textbf{Microscale Syn. Bias} & \textbf{Macroscale Syn. Bias} \\ \midrule
AND           & 1.623 bit               & 0.811 bit               & 0.533                             & 0.578                             \\
OR            & 2.811 bit               & 0.811 bit               & 0.518                             & 0.578                             \\
XOR           & 2.5 bit                 & 1 bit                   & 0.52                              & 0.833                             \\ \bottomrule
\end{tabular}
\label{tab:logic}
\caption{\textbf{The macro and microscale temporal mutual informations and their respective synergy biases.} For three logic gates (AND, OR, and XOR), this table shows the effects that going up a level of abstraction has on the temporal mutual informations and the synergy biases. It's important to understand that while the micro and macroscale implement the same function, the temporal mutual information can be very different.}
\end{table}

\begin{figure}
    \centering
    \includegraphics[scale=0.7]{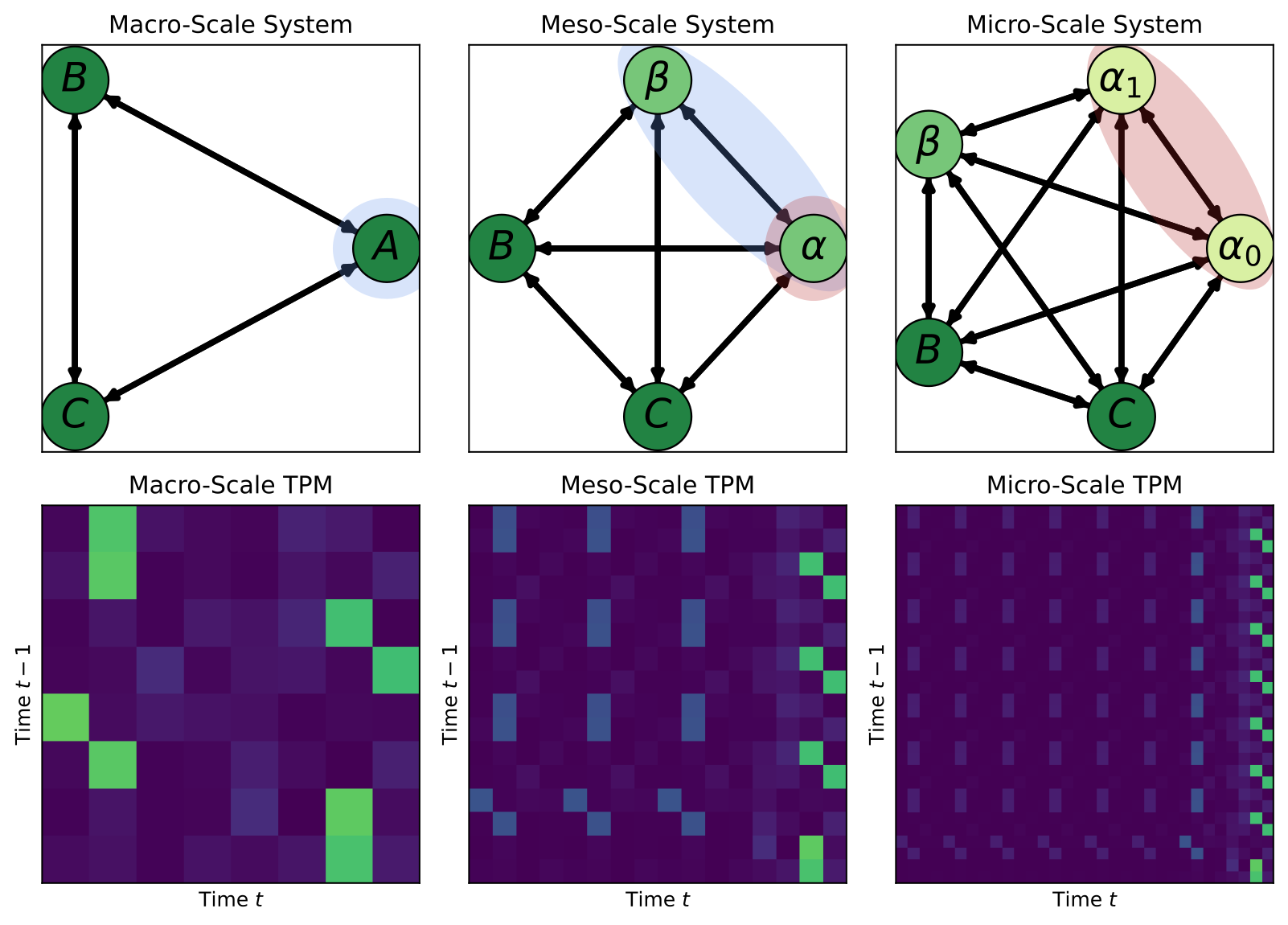}
    \caption{\textbf{Multi-scale analysis}. Here we can see how to construct equivalence class microscales from a given macroscale such that mutual information is fixed. \textbf{Left}: a three-element system (top), and its associated TPM. We select a single node ($A$) to expand. \textbf{Middle}: expanding node $A$ into nodes $\alpha$ and $\beta$ results in a four-element system, which crucially preserves the mutual information from the joint-past to the joint-future. We can select another node ($\alpha$) to expand again, resulting in \textbf{Right}: the final microscale expansion of our system. Note that the original node $A$ has been expanded twice, while the overall mutual information dynamics are preserved in all cases.}
    \label{fig:multi_scale}
\end{figure}

\subsection{\label{sec:modeling_across_scales}Redundant to synergistic information conversion}

In the cases of classic logic gate composition above in Section \ref{sec:logic_gates}, it could be argued by a skeptic that while the synergy biases are indeed increasing at the macroscale, this is because all the information at the bottom of the lattice is being removed by dimension reduction. This may be true in some instances. Luckily, we can provide direct evidence the effect we've observed is not just that some types of information (like redundant information at the bottom of the lattice) is lost at macroscales. Rather, there is evidence that information is being converted from one type to another (or more precisely, information is moving up the PI spectrum from redundant to more synergistic at the macroscale).

To show cases of information conversion, we developed a method by which the mutual information can be kept constant across scales. Since the mutual information in terms of total bits is not decreasing at the macroscale, any change in synergy bias must be from the conversion of information, not its loss.

First, it is important to note that a neglected aspect of fairly comparing micro and macroscales is making sure that the macroscales are viable models of the system. It is therefore critical that the dynamics between a macroscale model and a microscale model are either identical (as in our cases), or highly similar. That is, dimension reduction shouldn't lead to significant differences in dynamics, nor to responses to interventions, or else the macroscale is a poor model of the system. This has been called "consistency" and has been explored in structural equation modeling specifically of equivalence classes \citep{rubenstein2017causal}. Given that precise consistently is not always possible, it is possible to measure the amount of inconsistency as the difference between the dynamics of the microscale and that of the macroscale, and an informational measure of inconsistency was therefore introduced that can analyze how consistent a wide variety of systems are \citep{klein2020emergence}. It is worth noting that in this work, because of the use of equivalence classes in our method of expansion, in addition to the mutual information being constant, all scales used here also have zero inconsistency according to this measure. Such perfectly consistent macroscales do not necessarily need to be constructed via equivalence classes in order to ensure consistency, as we do here, since there are various types of macroscales that can give complete consistency \citep{klein2020emergence}.

What follows is the description of our method to hold the mutual information constant and ensure consistency between scales. This "expansion method" introduced here is based on generating a Boolean network (represented by some TPM). Assuming this Boolean network is a macroscale, we can then bifurcate nodes in the network in such a way to create an equivalence class: that is, \textit{n} elements that have the same function in that they have the same inputs and the same outputs. Effectively, this allows for the generation of microscales from a given macroscale, an inversion of the normal process of finding macroscales from a given microscale \citep{hoel2013quantifying}. Relevant Python code can be found in the supplementary materials attached to this manuscript. This process allows us to create different arbitrary microscales for a given system, while the mutual information remains fixed between the two scales. Therefore any change in bias on the PI spectrum comes from the conversion of information from one type to another.

To illustrate this, we constructed two hundred TPMs that were positive-Gaussian by initially generating $8\times8$ random Gaussian matrices from a distribution with a mean of 0 and a standard deviation of 1, taking the absolute value of every entry, and finally normalizing the rows to define discrete probability distributions. The resulting TPMs describe the stochastic dynamics of 200 distinct, fully connected three element systems with binary states. These are our starting macroscales. For each of these 200 systems, we split one element to create a 4-element system, two of which are from our initial macroscale, and two of which are ``children" of the initial split macroscale element. We then re-calculate the PI spectrum and synergy bias for our new microscale.  We can do this process of creating children in an equivalence class more than once: if so, we call the 4-node network a "mesoscale" and the 5-node network a "microscale." An example system is shown in Fig. \ref{fig:multi_scale}, as well as details of the entire process as it is expanded into a mesoscale and then microscale (non-macroscale nodes in Fig. \ref{fig:multi_scale} are referred to as $\alpha$ and $\beta$ in this process).

\begin{figure}
    \centering
    \includegraphics[scale=0.75]{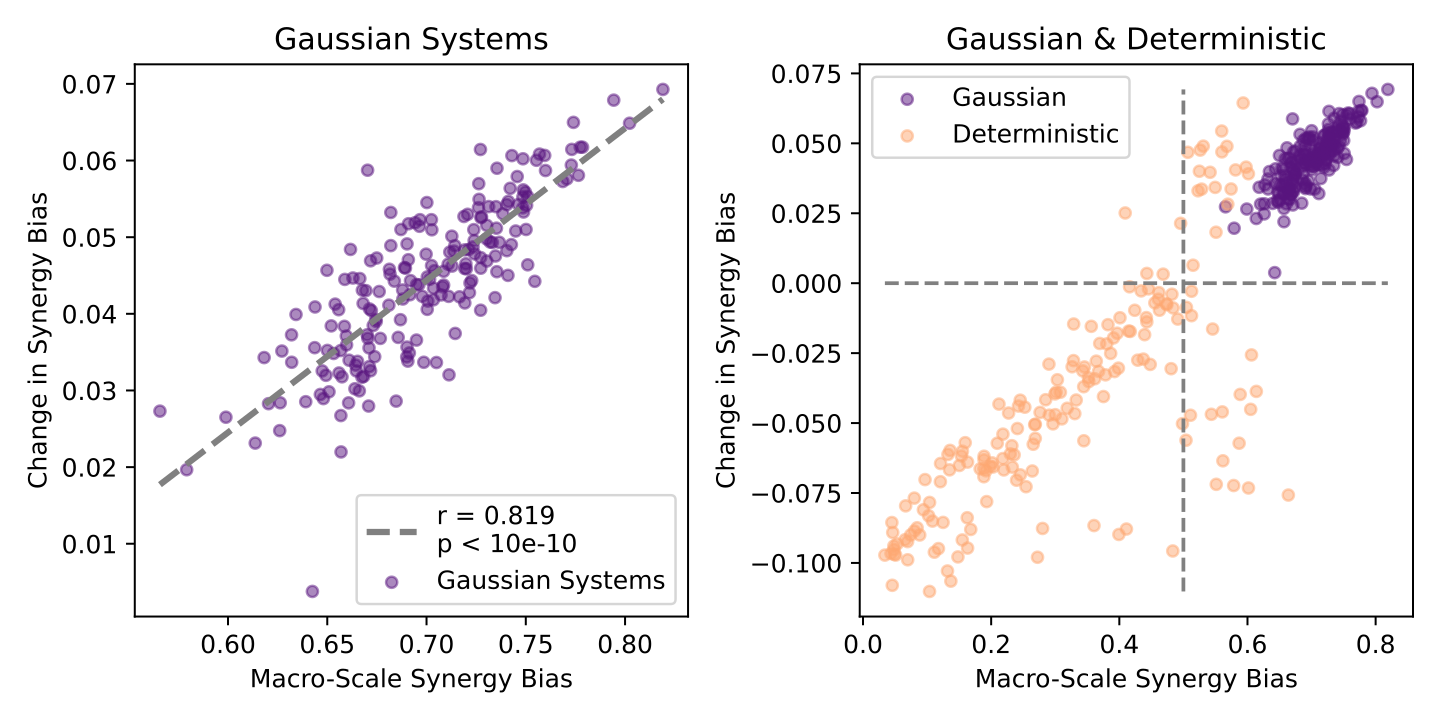}
    \caption{\textbf{Change in synergy bias across scales}. \textbf{Right:} the change in synergy from microscale to macroscale plotted against the starting macroscale synergy for Gaussian systems. There is a clear positive correlation. Indeed, all systems show an increase in synergy at the macroscale and an increase in redundancy at the microscale. The equivalence class structure used to construct the systems holds the mutual information steady across scales, so an increase in synergistic information at the macroscale can only come from a decrease in the redundant information of the microscale. \textbf{Left}: the same plot, although here we display both the Gaussian and deterministic systems. Note that, in contrast to the Gaussian systems, the deterministic systems start at a much lower synergy bias. Typically these lower synergy systems actually \textit{lose} synergy bias after coarse-graining, although the linear relationship between how synergistic the macroscale is and how much synergy is lost at the microscale remains. Interestingly, visual examination suggests that for both classes of system, the relationship between the macroscale synergy bias and the change in synergy bias is generally linear and, for both systems, lies along a common line of best fit. This suggests that, while different systems behave differently under coarse-graining, the broader relationship is preserved.}
    \label{fig:change_syn}
\end{figure}

First, all of the 200 Gaussian systems showed an increased bias toward synergy at the macroscale, despite the mutual information being unchanged across scales. In general, our hypothesis was that the higher the synergy bias of the macroscale, the more that synergy would be lost at the microscale. This appears to be true in these model systems; in Fig. \ref{fig:change_syn}, we correlated the gain in synergy bias following the conversion of the microscale to the macroscale, against the macroscale synergy bias. Pearson's product moment correlation found a highly significant positive correlation between macroscale synergy bias and the gain in synergy bias under repeated coarse grainings of the microscales (see Fig. \ref{fig:change_syn}, Left, $\rho = 0.819$ $p<10^{-10}$). This suggests that, for random stochastic systems, even when total mutual information is constant across scales, the systems have more redundant information at the microscale than the macroscale. This is proof that dimensionality reductions exist that increase the overall synergy of the system by converting information to be more synergistic.

In addition to the Gaussian matrices, we also constructed 185 ``deterministic" systems, where a single joint state lead predictably to another single joint state with probability 0.99 (the remaining probability was evenly distributed to ensure the system was ergodic, did not have fixed point attractor, and that the stationary distribution involved all system states), which we expanded in the same manner as described above into both ``meso" and microscales. By exploring both highly stochastic and deterministic systems, we can generate a richer sample of the space of all three-element Boolean systems and identify more universal patterns.

We found similar results with the deterministic systems, although several clear differences are immediately apparent upon inspection (see Fig \ref{fig:change_syn}, Right). First, it's clear that deterministic systems of the kind we are creating have a lower synergy bias overall than using the Gaussian method of creating systems. In cases where the macroscales are indeed synergistic, like they are in Gaussian systems (above 0.5 in synergy bias), then it is also the case that the underlying microscales are more redundant. However, in deterministic macroscales that are biased toward redundancy (below 0.5 in synergy bias) they can actually be expanded into microscales that are themselves more synergy biased (and the correlation between the change in synergy and the macroscale synergy remained positive).

\section{\label{sec:CE_as_information_conversion}Causal emergence as information conversion}

Given the evidence that information conversion can occur at macroscales, even in a well-understood baseline measure like the mutual information. How does it play out in measures like the effective information, which was specifically designed to capture causal influence and can peak at a macroscale?

A hint comes from the already well-established fact that the effective information ($EI$) changes across scales due to a change in determinism and/or degeneracy \citep{hoel2013quantifying, hoel2017map,klein2020emergence}. Indeed, it is already proven that:

\begin{equation}
\label{eq:mi}
    EI(X ; Y) = \sum_{x\in\mathcal{X}}\sum_{y\in\mathcal{Y}} P(x, y) \log_2\big(\frac{P(x,y)}{P(x)P(y)}\big)|P(X) = H^{max}
\end{equation}

which differs from the normal mutual information calculation in that $P(X)$ is set to $H^{max}$, and can be rewritten as 
\begin{equation}
\label{eq:ei}
    \text{Effective information} = \text{determinism} - \text{degeneracy}.
\end{equation}

In this interpretation of the effective information, the determinism is based on the information lost via uncertainty in state transitions:

\begin{equation} 
    \text{determinism} = \log_2(N) - \langle H(p(y)|P(X)=H^{max}) \rangle
\end{equation}

The term $\log_2(N)$ can be understood as the uncertainty about the future state of a maximally entropic system with $N$ unique states. The second term $\langle H(p(y)|P(X)=H^{max}) \rangle $ gives the average uncertainty about the future state of our real system $X$ (note that this is applied over a single timestep, e.g., $t$ to $t_{+1}$). The average uncertainty is a function of the noise, wherein $\langle H(p(y)|P(X)=H^{max}) \rangle $ is zero if there is no noise in any transition (and the system is therefore deterministic).

The difference between the two terms (the hypothetical maximum entropy and the empirical entropy) gives us a measure of how much how much better are we at predicting the future of $X$ than we would be in the ``worst case scenario." If effective information is increasing at a macroscale due to an increase in the determinism term, then the entropy term in the determinism must itself be necessarily decreasing. This is because $log2(N)$ also necessarily decreases at the macroscale, so therefore any increases in the determinism term must come from a greater decrease in $\langle H(p(y)|P(X)=H^{max}) \rangle $ than $log2(N)$. Figure 4 (left) shows this decrease in the information (the uncertainty of transitions), which can lead to an increase in effective information at a macroscale.

The degeneracy contains a similar entropy term and a size term:

\begin{equation} 
    \text{degeneracy} = \log_2(N) - H(\langle p(y)|P(X)=H^{max} \rangle)
\end{equation}

The degeneracy term is very similar structurally to the determinism term: once again the $\log_2(N)$ represents the maximally entropic ``reference" system. The term $H\langle (p(y)|P(X)=H^{max})\rangle$ quantifies the uncertainty in retrodiction of a previous state, given a current state. That is, degeneracy is the amount of information lost (in a single timestep) if the system is "reversed" in time.

The degeneracy can be thought of as the amount of information about the past that is lost when multiple causal paths ``run together." For instance, we can imagine a system where two unique states both lead to the same subsequent state. In this case, information about the past is lost because it is impossible to determine which of the two prior states preceded the current one from just the current state alone. It can also be thought of as a quantification of how different each state's transitions are. If in a system every state has a unique transition, then the degeneracy $H\langle( p(y)|P(X)=H^{max} \rangle)$ term is zero, and degeneracy maximal.

The case of degeneracy is more complicated, since here the entropy term can increase at macroscales. However, as in determinism, the $\log_2(N)$ term always decreases. And it is actually the decreasing of $\log_2(N)$ that can lead to a decrease in degeneracy at the macroscale. E.g., if two states both deterministically transition to a single state, then a grouping over those (or an equivalent drop out of one) leads to an decrease in in degeneracy since $\log_2(N)$ decreases. This leads to an increase in $EI$ because the degeneracy term itself is subtracted in the $EI$ equation.

Notably, it is well-known that the effective information cannot increase at a macroscale if determinism is maximal and degeneracy is minimal. Why? Because there is no information to convert, neither in terms of the size of the system $\log_2(N)$ nor in terms of the uncertainty of transitions $\langle H(p(y)|P(X)=H^{max}) \rangle $. Therefore, causal emergence can be conceived as the conversion of causally-irrelevant information (like the uncertainty of state transitions) to causally-relevant information (the effective information), fitting into the umbrella theory of emergence as information conversion. 


\section{\label{sec:discussion}Discussion}

We have offered forth an umbrella theory of emergence based on how changes in scale lead to the conversion of one type of information to another. That is, dimension reduction does not necessarily leave all types of information invariant. We claim that the best way to consider these questions is to examine how changes in modeling, such as dimension reduction, change information type. While some information measures, like the total correlation between past and future measured by the mutual information, can only stay constant or decrease at macroscales, such measures can still demonstrate information conversion (such as here from redundant to synergistic information). We have shown this effect in Boolean networks of logic gates, such as comparing a macroscale XOR to its underlying logic gates, none of which are XORs at the microscale. We have also shown it in cases of equivalence classes where the mutual information is held constant across scales and yet still information can become more biased toward synergy at the macroscale, proving information conversion.

Further future work may be examining things like at what scale synergistic information peaks, or how to find scales that maximally convert information while minimally losing information. Though we have shown evidence that some redundant information must become synergistic (or vice versa), it remains to be understood exactly \textit{which} information changes form. Recent work on decomposing the local mutual information into directed local entropies may provide an interesting path forward \citep{finn_probability_2018}. Another promising future direction of research would be to introduce local information analysis to this pipeline \citep{lizier_local_2013,finn_pointwise_2018}. 

More broadly, this umbrella theory of emergence reveals that there are measurable benefits to macroscale models. If so, this is likely to have been selected for in science, i.e., members of the special sciences can be viewed as converting redundant information into a more useful form. Indeed, it has even been shown that synergistic information processing can be key to certain games, like to a successful poker strategy \citep{frey2018synergistic}. Our hypothesis is that the special sciences, and macroscale models in general, involve the conversion of redundant information into synergistic and unique information, making such macroscales useful for experimenters above and beyond their degree of compression. 

Tying this to previous research, macroscale modeling can also convert causally-irrelevant information to causally-relevant information by making causal relationships between variables in a model more dependent (by increasing determinism or decreasing degeneracy), i.e., causal emergence.
Note that there are clear advantages to identifying scales at which variables are more dependent. For instance, it has been shown that biological networks show more causal emergence than comparable technological or social networks \citep{klein2020emergence}. This is probably because there are multiple advantages for macroscales, ranging from a lower entropy of random walkers to greater global efficiency at macroscales \citep{griebenow2019finding}. Some preliminary research examining the protein interactomes of over 1,000 species shows that macroscales have become more likely to demonstrate causal emergence over evolutionary time \citep{hoel2020evolution}. This may even be one reason that controlling biological systems is so difficult: they are cryptic by having an intrinsic functional scale be a difficult-to-discover macroscale, making biological networks more robust to failure and less likely to be controlled from the outside \citep{hoel2020emergence}. 

The sort of above analyses are just a fraction of the applications of a formal theory of emergence, which is ultimately a toolkit for identifying the intrinsic scale of function of complex systems. This issue of identifying intrinsic scale crops up all across the sciences, such as modeling gene regulatory networks in biology \citep{hoel2020emergence}, understanding whether the brain functions at the scale of neurons or minicolumns \citep{buxhoeveden2002minicolumn, yuste2015neuron}, answering what level of abstraction is appropriate for modeling and comparing deep neural networks \citep{cao2021explanatory}, or even examining the best scale to model cardiac systems at \citep{ashikaga2018causal}. Previous research has shown how intrinsic scale comes about via growth rules, e.g., networks that develop causally-emergent macroscales only do so once the network is no longer growing in a "scale-free" manner \citep{klein2020emergence}. Ultimately, by tracking information conversion, experimenters and modelers can close in on the intrinsic scale of function for a given system.

\subsection{\label{sec:comparison_to_others}Comparison to other theories of emergence}

To help the nascent field of formal theories of emergence, it is important to discuss exactly what type of emergence we mean here, and compare and contrast to other definitions. For example, there is no doubt that simple laws and interactions in systems can lead to the emergence of complex, interesting, or unexpected dynamics, such as in cases of symmetry breaking \citep{anderson1972more} or simple rule iteration \citep{wolfram2002new}. Sometimes this is referred to as "emergence." However, this phenomena of complexity emerging from simplicity is not conceptually mysterious, and is already well-understood mathematically.

Another use of the term "emergence" comes from thinking about joint effects, which is a "whole vs. parts" emergence. Ultimately, this is motivated by the fact that elements in a system can exhibit behavior, dynamics, or functions that would not take place if they were partitioned or isolated from the rest of the system. One such measure that captures how much information is lost by partitioning individual elements off from a given system is Integrated Information Theory \citep{tononi2008consciousness, oizumi2014phenomenology}. A similar proposition is the ``integrated information decomposition" ($\Phi$ID) \citep{mediano_beyond_2019,rosas_reconciling_2020}. As with the work described here, the $\Phi$ID framework actually takes as its starting point the decomposition of the mutual information between past and future. However, the $\Phi$ID framework constructs a double PI lattice that describes how information moves from one PI-atom to another through time. In this framework, the authors define ``emergent" information as information that remains synergistically present across the joint state of many elements through time. They also refer to information in the joint state of the whole that constrains the evolution of a particular element as ``downward causation." However, as in IIT both of these are just joint information flow over sets of elements at a single scale, and there is an absence of any kind of macroscale vs. microscale comparison. It's worth noting that sets of elements with joint effects may be good candidates for macroscales. However, the mere fact that joint sets of elements behave differently compared to isolated elements in terms of effects or information flow does not by itself capture what is lost by reduction to some lower scale of explanation.

The "macro vs. micro emergence" kind discussed here resembles that which is traditionally discussed in analytic philosophy, which involves issues of supervenience, multiple-realizablity, and causality \citep{humphreys2016emergence}. There it is sometimes called "synchronic" emergence \citep{humphreys2008synchronic}, although we eschew this term as confusing for scientific usage. However, as we have shown, a mathematical theory of emergence is not put forward to resolve metaphysical problems, but is part of modeler and experimenter toolkits when it comes to identifying intrinsic scale of function, as well as modeling practices. This process involves explicit modeling of different scales of model or physical systems, followed by their comparison. While related to philosophical discussions of emergence, note that our proposal of emergence as information conversion does not fit cleanly into the traditional strong/weak emergence dichotomies in philosophy \citep{chalmers2006strong}. Supervenience is not violated when information is converted from one type to another (such as redundant mutual information becoming synergistic, uncertainty of transitions being transformed into effective information, etc). In this view, reduction is always possible when supervenience holds, and therefore there is always an identifiable procedure to map one scale to another. However, such reduction can lead to a real and measurable loss of a given type of information. This offers a subtle but powerful explanation as to what advantages macroscale models (such as the existence of the special sciences), provide above and beyond compression.

\section{\label{sec:acknowledgments}Acknowledgments}

This work was supported and made possible by the USA Army Research Office (proposal 77111-PH-II) and the NSF-NRT grant 1735095, Interdisciplinary Training in Complex Networks and Systems at Indiana University Bloomington. 

\bibliographystyle{plainnat}

\bibliography{biblio}

\end{document}